\documentclass[aps,prb,reprint,twocolumn,amsmath,amssymb]{revtex4-2}
\usepackage{stmaryrd}
\usepackage{times}
\usepackage{graphicx}
\usepackage{dcolumn}
\usepackage{bm}
\usepackage{color}
\usepackage{makecell}
\usepackage{tabularx,multirow}
\usepackage{amsmath}
\usepackage{amssymb}% http://ctan.org/pkg/amssymb
\usepackage{pifont}% http://ctan.org/pkg/pifont
\usepackage{float}
\newcommand{\cmark}{\ding{51}}%
\newcommand{\xmark}{\ding{55}}%
\begin{document}
\title{Anisotropic spin-polarized conductivity in collinear altermagnets}
\author{Mingbo Dou$^{1}$, Xianjie Wang$^{1,2}$ and L. L. Tao$^{1,2}$}
\email{Contact author: lltao@hit.edu.cn}
\affiliation{$^{1}$ School of Physics, Harbin Institute of Technology, Harbin 150001, China}
\affiliation{$^{2}$ Heilongjiang Provincial Key Laboratory of Advanced Quantum Functional Materials and Sensor Devices, Harbin 150001, China}
\date{\today}
\begin{abstract}
The altermagnet exhibits the nonrelativistic spin splitting that enables all-electrical generation of spin-polarized currents beyond the spin-orbit coupling. Here, we report on a study on the anisotropic spin-polarized conductivity in collinear altermagnets. Based on the Boltzmann transport theory, we first study this effect using the general group-theoretical analysis and identify the spin point groups sustaining the finite spin polarization defined in terms of spin-polarized conductivity. We show that the spin polarization vanishes along any direction for the $g$-wave and $i$-wave altermagnets while the spin polarization is significantly anisotropic for the $d$-wave altermagnet. We further derive the analytical expressions for the anisotropic spin polarization in the $d$-wave altermagnets. Then, we exemplify those phenomena in several representative altermagnets based on the density functional theory calculations. Our work enriches the altermagnetic spintronics and paves the practical way to produce large spin polarization in collinear altermagnets.
\end{abstract}
\maketitle
\section{Introduction}
Recently, the altermagnet characterized by the zero net magnetization and nonrelativistic momentum-dependent spin splitting has been proposed based on the spin group symmetry\cite{prx031042,prx040501,prx021016,prx031037,prx031038,prx031039}. It represents a third distinct magnetic phase in the sense that it reveals both ferromagnetic (nonrelativistic spin splitting\cite{jpsj123702,prb014422}) and antiferromagnetic (zero magnetization) characteristics. Thus, the advantages of antiferromagnets\cite{rmp015005} such as vanishing stray fields and high-frequency spin dynamics are inherent in altermagnets. Moreover, the altermagnet exhibits the $\mathcal{T}$-symmetry-breaking ($\mathcal{T}$ for time reversal) anomaly responses and spin-orbit effects beyond the spin-orbit coupling (SOC)\cite{na509,jpd113001,prl076801} such as anomalous Hall effect\cite{eaaz8809,ne735}, charge to spin conversion\cite{prl127701}, spin-splitter torque\cite{prl127701,prl137201}, and giant and tunneling magnetoresistance effects\cite{prx011028,nc7061}, etc. Those combined intriguing properties make the altermagnet suitable to explore the novel physical phenomenon and promising for various spintronics applications\cite{afm2409327,jcas2257,nrm2025}.

One of the intriguing properties offered by altermagnets is the spin-polarized conductivity, which is forbidden in conventional collinear antiferromagnets due to the spin degeneracy of energy bands. This is legitimate since the group velocities are spin-dependent due to the nonrelativistic spin splitting, that is, $v^\uparrow_i\neq v^\downarrow_i$ ($i$ for direction and $\uparrow, \downarrow$ for spins), as pointed out and demonstrated in the RuO$_2$ by \ifmmode \check{S}\else \v{S}\fi{}mejkal \emph{et al}\cite{prx011028}. In an earlier study, Gonz\'alez-Hern\'andez \emph{et al} \cite{prl127701} demonstrated the highly efficient spin-current generation in the RuO$_2$ and developed the framework for the Laue-group symmetry dictated spin-conductivity tensor. However, the complete spin-point-group (SPG) dictated spin-polarized conductivity, and in particular its spatial anisotropy remain largely unexplored. Here, we investigate the spin-polarized conductivity for collinear altermagnets using the Boltzmann transport theory. We discuss the spin-polarized conductivity anisotropy and identify the SPGs sustaining the finite spin polarization. Moreover, we calculate the spin-polarized conductivity and anisotropic spin polarization for several representative collinear altermagnets based on the density functional theory (DFT) calculations.

The rest of the paper is organized as follows. In Sec. II, we present the theoretical formalism and computational details for spin-polarized conductivity calculations. In Sec. III, we discuss the spin-polarized conductivity anisotropy based on the general symmetry analysis. In Sec. IV, we present the DFT results for several representative altermagnets. Finally, Sec. V is reserved for further discussion and conclusion.

\section{Theoretical formalism and computational details}

\begin{table*}
\caption{\label{table1} Symmetry allowed (\cmark, $\sigma^\uparrow_{ij}\neq\sigma^\downarrow_{ij}$), forbidden (\xmark, $\sigma^\uparrow_{ij}=\sigma^\downarrow_{ij}$) spin-polarized conductivity and zero conductivity ($\sigma^\uparrow_{ij}=\sigma^\downarrow_{ij}=0$) for different SPGs sustaining the collinear altermagnets. $SP_\mathbf{\hat{n}}$ represents the anisotropic spin polarization defined by Eq. \ref{eq-7} in the main text.}
\begin{ruledtabular}
\begin{tabular}{cccccccccc}
SPG & $\sigma_{xx}$ & $\sigma_{yy}$ & $\sigma_{zz}$ & $\sigma_{xy}$ & $\sigma_{yz}$ & $\sigma_{xz}$ & $SP_\mathbf{\hat{n}}$ & Candidate & Anisotropy\\
\hline
$^2m^2m^1m$ & \xmark & \xmark & \xmark & \cmark & $0$ & $0$ & $\frac{2\sigma_{xy}^\uparrow\text{sin}^2\theta\text{sin}(2\varphi)}{\sigma_{xx}\text{sin}^2\theta\text{cos}^2\varphi+\sigma_{yy}\text{sin}^2\theta\text{sin}^2\varphi+\sigma_{zz}\text{cos}^2\theta}$ & FeSb$_2$ & $d$-wave\\
$^24/^{1}m$ & \cmark & \cmark & \xmark & \cmark & $0$ & $0$ & $\frac{(\sigma_{xx}^\uparrow-\sigma_{xx}^\downarrow)\text{sin}^2\theta\text{cos}(2\varphi)+2\sigma_{xy}^\uparrow\text{sin}^2\theta\text{sin}(2\varphi)}{\sigma_{xx}\text{sin}^2\theta+\sigma_{zz}\text{cos}^2\theta}$ & KRu$_4$O$_8$ & $d$-wave\\
$^24/^1m^2m^1m$ & \xmark & \xmark & \xmark & \cmark & $0$ & $0$ & $\frac{2\sigma_{xy}^\uparrow\text{sin}^2\theta\text{sin}(2\varphi)}{\sigma_{xx}\text{sin}^2\theta+\sigma_{zz}\text{cos}^2\theta}$ &  RuO$_2$ & $d$-wave\\
$^14/^1m^2m^2m$ & \xmark & \xmark & \xmark & $0$ & $0$ & $0$ & 0 & KMnF$_3$ & $g$-wave\\
$^16/^1m^2m^2m$ & \xmark & \xmark & \xmark & $0$ & $0$ & $0$ & 0 & $-$ & $i$-wave\\
$^22/^2m$ & \xmark & \xmark & \xmark & \cmark & \cmark & \xmark & $\frac{2\sigma_{xy}^\uparrow\text{sin}^2\theta\text{sin}(2\varphi)+2\sigma^\uparrow_{yz}\text{sin}(2\theta)\text{sin}\varphi}{\sigma_{xx}\text{sin}^2\theta\text{cos}^2\varphi+\sigma_{yy}\text{sin}^2\theta\text{sin}^2\varphi+\sigma_{zz}\text{cos}^2\theta+\sigma_{xz}\text{sin}(2\theta)\text{cos}\varphi}$ & CuF$_2$ & $d$-wave\\
$^1\bar{3}^2m$ & \xmark & \xmark & \xmark & $0$ & $0$ & $0$ & 0 & CoF$_3$ & $g$-wave\\
$^26/^2m$ & \xmark & \xmark & \xmark & $0$ & $0$ & $0$ & 0 & $-$ & $g$-wave\\
$^26/^2m^2m^1m$ & \xmark & \xmark & \xmark & $0$ & $0$ & 0 & 0 & CrSb & $g$-wave\\
$^1m^1\bar{3}^2m$ & \xmark & \xmark & \xmark & $0$ & $0$ & $0$ & 0 & $-$ & $i$-wave\\
\end{tabular}
\end{ruledtabular}
\end{table*}

Assume that a direct electric field $\mathbf{\mathcal{E}}$ is applied to a solid and $\mathbf{J}$ is the produced current density. To first order in $\mathbf{\mathcal{E}}$, the spin-resolved $\mathbf{J}^s$ ($s=\uparrow, \downarrow=\pm1$ for spin index) is found as\cite{callaway}
\begin{equation}\label{eq-1}
  J_i^s=\sigma_{ij}^s\mathcal{E}_j,
\end{equation}
where $\sigma_{ij}^s$ is the conductivity tensor, the indices $i, j$ denote Cartesian components and a summation over repeated indices is implied. Under the relaxation time $\tau$ approximation, $\sigma_{ij}^s$ reads\cite{callaway}
\begin{equation}\label{eq-2}
    \sigma _{ij}^{s}(\epsilon_F) = -\frac{e^2\tau}{8\pi^3\hbar^2}\sum_n\int\frac{\partial\epsilon_{n\mathbf{k}}^s}{\partial k_i}\frac{\partial\epsilon_{n\mathbf{k}}^s}{\partial k_j}\frac{\partial f^0}{\partial\epsilon_{n\mathbf{k}}^s}d^3k,
\end{equation}
where $f^0(\epsilon_{n\mathbf{k}}^s, \epsilon_F)$ is the Fermi distribution function given in terms of the eigenvalue of the $n$th band $\epsilon_{n\mathbf{k}}^s$ and the Fermi energy $\epsilon_F$, $\hbar$ the reduced Planck's constant, and $\mathbf{k}=(k_x, k_y, k_z)$ the wave vector given in the Cartesian coordinates. It is evident that the transformation of $\sigma_{ij}^s$ under symmetry
operations is equivalent to that of $sk_ik_j$, which can thus be used as a symmetry equivalent quantity. $\sigma_{ij}^s$ is spin polarized (degenerate) if $\sigma_{ij}^\uparrow\neq\sigma_{ij}^\downarrow$ ($\sigma_{ij}^\uparrow=\sigma_{ij}^\downarrow$). In addition, $\sigma_{ij}^s$ is symmetry forbidden if $\sigma_{ij}^s\rightarrow-\sigma_{ij}^s$ under a chosen operation, e.g., rotation or mirror reflection.

According to previous work\cite{prx031042}, the collinear altermagnet is described by the third type of SPG $\mathbf{R}^{\text{III}}_{s}$,
\begin{equation}\label{eq-3}
    \mathbf{R}^{\text{III}}_{s}=[E||\mathbf{H}]+[C_2||A][E||\mathbf{H}]=[E||\mathbf{H}]+[C_2||\mathbf{G-H}],
\end{equation}
where $\mathbf{G}$ is the crystallographic Laue group, $\mathbf{H}$ is the halving subgroup of $\mathbf{G}$ and contains the spatial inversion element, $A$ represents the real-space rotation and satisfies $\mathbf{G-H}=A\mathbf{H}$. $E$ and $C_2$ are the identity and $180^\text{o}$ rotation around an axis perpendicular to the spin, respectively. Let $R$ being the symmetry element of $\mathbf{G}$, $\sigma^s$ due to the symmetry constraints satisfies
\begin{equation}\label{eq-4}
  \mathbf{D}(R)\sigma^{\{E/C_2\}s}\mathbf{D}(R)^\dag=\sigma^s,
\end{equation}
where $\mathbf{D}(R)$ is the representation matrix of $R$ operation. For a given SPG, the spin-polarized conductivity component $\sigma^s_{ij}$ can be deduced from  Eq. \ref{eq-4}.

Our DFT calculations were performed using the plane-wave ultrasoft pseudopotential method\cite{prb7892} as implemented in the QUANTUM ESPRESSO\cite{jpcm395502,jpcm465901,jcp154105}. An energy cutoff of $50$ Ry for the plane-wave expansion and generalized gradient approximation (GGA)\cite{prl3865} for the exchange and correlation functional were adopted throughout. The other computational parameters such as $k$-point mesh for the Brillouin zone integration, lattice parameters and Hubbard-$U$ values shall be detailed for the corresponding materials.

The spin-resolved conductivity $\sigma_{ij}^s$ was calculated by using the Boltzmann transport theory under relaxation time $\tau$ approximation (Eq. \ref{eq-2}), as implemented in the BoltzWann module\cite{BoltzWann} of the Wannier90 code\cite{jpcm165902}, which is based on the maximally-localized Wannier function basis set\cite{prb12847,rmp1419}. We used the temperature of $300$ K in the Fermi distribution function and the $k$-point mesh for conductivity calculations  shall be detailed for the corresponding materials.

\section{Group-theoretical analysis}

\begin{figure*}
\includegraphics[width=0.8\textwidth]{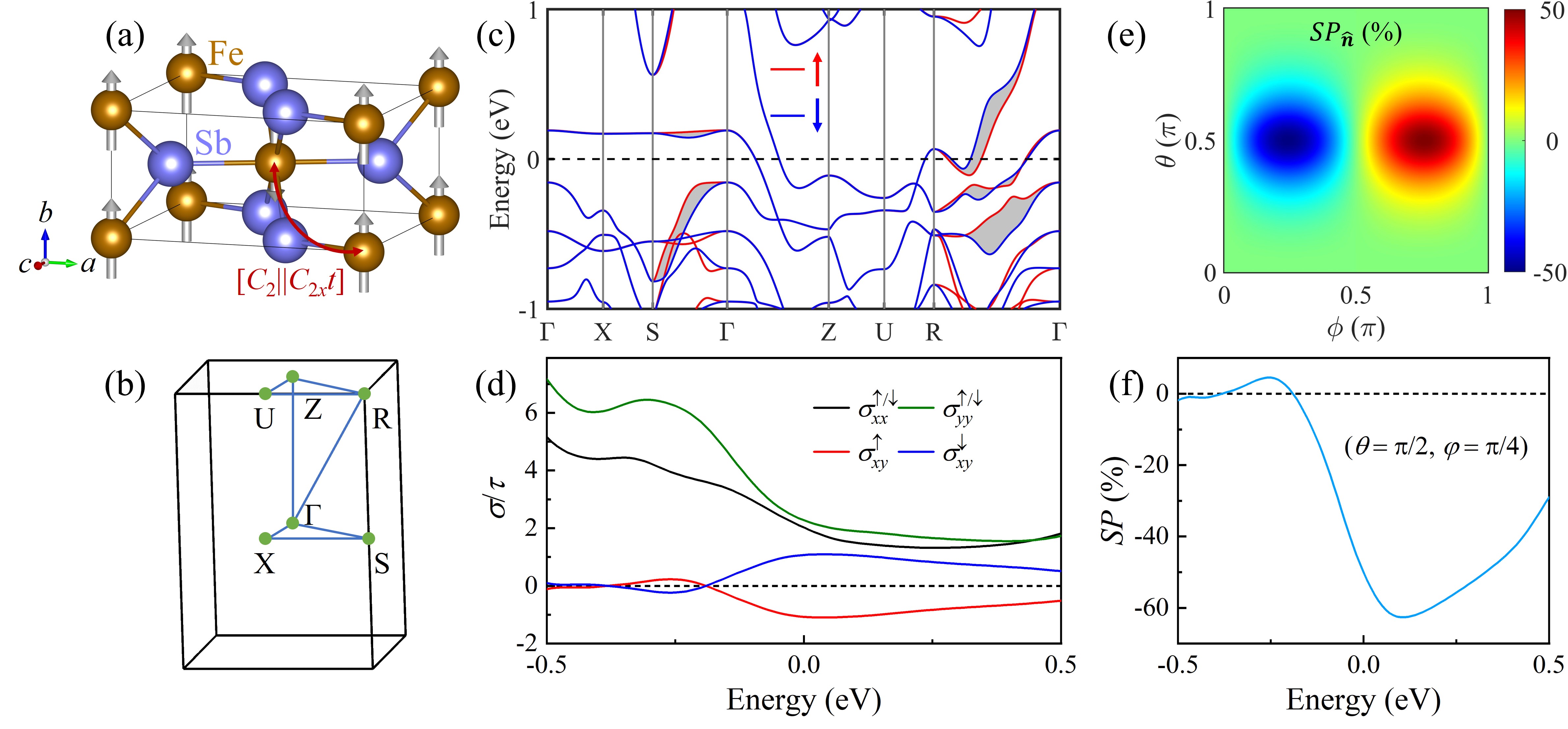}%
\caption{\label{f-1} (a) Crystal structure and (b) Brillouin zone of the orthorhombic FeSb$_{2}$ (space group $Pnnm$). In (a), the gray arrows denote the magnetic moments of Fe atoms and the opposite-spin sublattices are connected by the symmetry $[C_2||C_{2x}t]$ with the translation $t=(1/2, 1/2, 1/2)$. (c) Spin-resolved band structure (Red for spin up and blue for spin down) without SOC along the $k$-space path: $\Gamma(0, 0, 0)-X(0.5, 0, 0)-S(0.5, 0.5, 0)-\Gamma-Z(0, 0, 0.5)-U(0.5, 0, 0.5)-R(0.5, 0.5, 0.5)-\Gamma$ indicated in (b). The gray colored regions highlight the spin splitting. (d) Spin-resolved conductivity (unit: $10^{19}$ $\Omega^{-1}$m$^{-1}$s$^{-1}$) as a function of energy. The Fermi energy has been aligned to zero. (e) Spin polarization $SP_\mathbf{\hat{n}}$ at the Fermi energy as a function of ($\theta$, $\varphi$). (f) Spin polarization $SP$ at ($\theta=\pi/2$, $\varphi=\pi/4$) as a function of energy.}
\end{figure*}

We first analyze the spin-polarized conductivity using the symmetry analysis presented above. As a illustration, we consider the SPG $^2m^2m^1m$ and the similar arguments can be applied to other SPGs readily. As demonstrated previously\cite{prx031042}, the SPG $^2m^2m^1m$ can be expressed as
\begin{equation}\label{eq-5}
\begin{aligned}
    ^2m^2m^1m&=[E||2/m]+[C_2||C_{2x}][E||2/m]\\
    &=\{[E||E],[E||I],[E||C_{2z}],[E||M_{z}]\}+\\
    &\{[C_2||C_{2x}],[C_2||C_{2y}],[C_2||M_{x}],[C_2||M_{y}]\},
\end{aligned}
\end{equation}
where $I$ is the spatial inversion, $C_{2i}$ ($i=x, y, z$) is the twofold rotation around the $i$-axis, and $M_i$ is mirror reflection about the $i=0$ plane. According to the transformation properties of $sk_ik_j$, the $[E||M_{z}]$ operation enforces that $\sigma_{xz}^s=\sigma_{yz}^s=0$ while the $[C_2||C_{2i}]$ or $[C_2||M_{i}]$ ($i=x, y$) operation yields $\sigma_{xx}^\uparrow=\sigma_{xx}^\downarrow$, $\sigma_{yy}^\uparrow=\sigma_{yy}^\downarrow$, $\sigma_{zz}^\uparrow=\sigma_{zz}^\downarrow$ and $\sigma_{xy}^\uparrow=-\sigma_{xy}^\downarrow$. It is indicated here that only $\sigma_{xy}$ is spin polarized.

A similar argument applies to other SPGs and Table \ref{table1} summarizes the symmetry-allowed (\cmark) or forbidden (\xmark) spin-polarized $\sigma^s_{ij}$ for ten different SPGs sustaining the collinear altermagnet. We first examine the non-diagonal components $\sigma^s_{ij}$ ($i\neq j$). If the subgroup $\mathbf{H}$ in Eq. \ref{eq-3} contains the rotation operation $C_n$ with $n\geq3$, all the non-diagonal components are null, that is, $\sigma^s_{xy}=\sigma^s_{yz}=\sigma^s_{xz}=0$. This is observed in the SPGs $^14/^1m^2m^2m$ ($C_{4z}$), $^16/^1m^2m^2m$ ($C_{6z}$), $^1\bar{3}^2m$ ($C_{3z}$), $^26/^2m$ ($C_{3z}$), $^26/^2m^2m^1m$ ($C_{3z}$), and $^1m^1\bar{3}^2m$ ($C_{3}$). In addition, the symmetry elements $[C_2||C_{2x}]$ ($^2m^2m^1m$) and $[C_2||C_{4z}]$ ($^24/^{1}m$ and $^24/^1m^2m^1m$) lead to spin-polarized $\sigma_{xy}$, that is, $\sigma^\uparrow_{xy}=-\sigma^\downarrow_{xy}$. In the case of $^22/^2m$, the symmetry element $[C_2||M_{y}]$ gives rise to the spin-polarized $\sigma_{xy}$ and $\sigma_{yz}$, that is, $\sigma^\uparrow_{xy}=-\sigma^\downarrow_{xy}$ and $\sigma^\uparrow_{yz}=-\sigma^\downarrow_{yz}$. On comparison, we see that only the SPG $^24/^{1}m$ sustains the spin-polarized diagonal components $\sigma_{xx}$ and $\sigma_{yy}$. Moreover, the symmetry element $[C_2||C_{4z}]$ yields $\sigma^\uparrow_{xx}=-\sigma^\downarrow_{yy}$ and $\sigma^\downarrow_{xx}=-\sigma^\uparrow_{yy}$.

It is enlightening to examine the spin-polarized conductivity $\sigma^s_\mathbf{\hat{n}\hat{n}}$ along the direction $\mathbf{\hat{n}}=(\text{sin}\theta\text{cos}\varphi, \text{sin}\theta\text{sin}\varphi, \text{cos}\theta)$ ($\theta$ for polar angle and $\varphi$ for azimuthal angle). With the aid of directional derivatives, one finds
\begin{equation}\label{eq-6}
\begin{aligned}
    &\sigma^s_\mathbf{\hat{n}\hat{n}}=\sigma^s_{xx}\text{sin}^2\theta\text{cos}^2\varphi+\sigma^s_{yy}\text{sin}^2\theta\text{sin}^2\varphi+\sigma^s_{zz}\text{cos}^2\theta\\
    &+\sigma^s_{xy}\text{sin}^2\theta\text{sin}(2\varphi)+\sigma^s_{yz}\text{sin}(2\theta)\text{sin}\varphi+\sigma^s_{xz}\text{sin}(2\theta)\text{cos}\varphi.
\end{aligned}
\end{equation}
Then, one can define the anisotropic spin polarization $SP_\mathbf{\hat{n}}$ as
\begin{equation}\label{eq-7}
  SP_\mathbf{\hat{n}}=\frac{\sigma^\uparrow_\mathbf{\hat{n}\hat{n}}-\sigma^\downarrow_\mathbf{\hat{n}\hat{n}}}{\sigma^\uparrow_\mathbf{\hat{n}\hat{n}}+\sigma^\downarrow_\mathbf{\hat{n}\hat{n}}}.
\end{equation}

In Table \ref{table1}, we summarize $SP_\mathbf{\hat{n}}$ for different SPGs. We first note that $SP_\mathbf{\hat{n}}$ is always zero for the SPGs $^14/^1m^2m^2m$, $^16/^1m^2m^2m$, $^1\bar{3}^2m$, $^26/^2m$, $^26/^2m^2m^1m$ and $^1m^1\bar{3}^2m$ although the nonrelativistic spin splitting occurs in those collinear altermagnets. This leads to the recognition that spin splitting is not the sufficient condition for the spin-polarized conductivity in collinear altermagnets. Moreover, our result expresses the fact that the conductivity for the $g$-wave and $i$-wave altermagnets is always spin degenerate while the $d$-wave altermagnet can sustain the spin-polarized conductivity along certain directions.

\section{DFT results}

Having established the SPG dictated spin-polarized conductivity and anisotropic spin polarization based on the general symmetry analysis, we next exemplify those phenomenon in several representative collinear altermagnets based on the DFT calculations. It is noteworthy that the SPG and band splitting for those altermagnets have been studied by previous work. We shall recall some important spin and electronic structures and focus on the anisotropic spin-polarized conductivity.

\subsection{FeSb$_2$}

FeSb$_2$ crystallizes in a orthorhombic structure with the space group $Pnnm$ shown in Fig. \ref{f-1}(a) and Fig. \ref{f-1}(b) shows its Brillouin zone with high-symmetry $k$ points indicated. We infer from Fig. \ref{f-1}(a) that the opposite-spin sublattices are connected by the symmetry $[C_2||C_{2x}t]$ with the translation $t=(1/2, 1/2, 1/2)$ and the corresponding SPG is $^2m^2m^1m$. We used the experimental lattice constants\cite{pnase2108924118} $a=5.834$, $b=6.53$ and $c=3.193$ {\AA} and Monkhorst-Pack mesh of $8\times8\times16$ for DFT calculations. Figure \ref{f-1}(c) shows the spin-resolved band structure without SOC. One observes that the sizable band splitting occurs along the $\Gamma-S$ and $\Gamma-R$ paths consistent with previous work\cite{pnase2108924118}. This is due to the fact that the $k$ points along those $k$ paths are not invariant under the symmetry operations in the subgroup $\mathbf{G}-\mathbf{H}=\{C_{2x}, C_{2x}, M_x, M_y\}$.

Figure \ref{f-1}(d) shows the spin-resolved conductivity as a function of energy calculated by using the $k$-point mesh of $110\times110\times200$. It is seen that $\sigma_{xx}$ and $\sigma_{yy}$ (also $\sigma_{zz}$, not shown) are spin-degenerate while $\sigma_{xy}$ is spin polarized characterized by $\sigma^\uparrow_{xy}=-\sigma^\downarrow_{xy}$ as consistent with the above SPG symmetry analysis (see Table \ref{table1}). From Eqs. \ref{eq-6} and \ref{eq-7} and Table \ref{table1}, the anisotropic spin polarization $SP_\mathbf{\hat{n}}$ can be calculated as
\begin{equation}\label{eq-8}
  SP_\mathbf{\hat{n}}=\frac{2\sigma_{xy}^\uparrow\text{sin}^2\theta\text{sin}(2\varphi)}{\sigma_{xx}\text{sin}^2\theta\text{cos}^2\varphi+\sigma_{yy}\text{sin}^2\theta\text{sin}^2\varphi+\sigma_{zz}\text{cos}^2\theta}.
\end{equation}
From Eq. \ref{eq-8}, $SP_\mathbf{\hat{n}}$ reveals a period $\pi$ for $\varphi$ and the extrema of $SP_\mathbf{\hat{n}}$ are at $\theta=\pi/2$. Figure \ref{f-1}(e) shows $SP_\mathbf{\hat{n}}$ at the Fermi energy as a function of ($\theta$, $\varphi$). The overall trend is that $SP_\mathbf{\hat{n}}$ is significantly anisotropic characterized by the sizable or zero spin polarization for certain ($\theta$'s, $\varphi$'s) and the spin polarization could reach as high as $50\%$. It is also seen that the magnitude of $SP_\mathbf{\hat{n}}$ is sizable around ($\theta=\pi/2$, $\varphi=\pi/4$) and ($\theta=\pi/2$, $3\pi/4$). This is legitimate since only the in-plane component $\sigma_{xy}$ is spin polarized. Figure \ref{f-1}(f) shows $SP_\mathbf{\hat{n}}$ at ($\theta=\pi/2$, $\varphi=\pi/4$) as a function of energy. It is observed that the magnitude of spin polarization reaches the maximum around $0.1$ eV and maintains $\sim50\%$ around the Fermi energy.

\subsection{K$_2$Ru$_8$O$_{16}$}

\begin{figure*}
\includegraphics[width=0.8\textwidth]{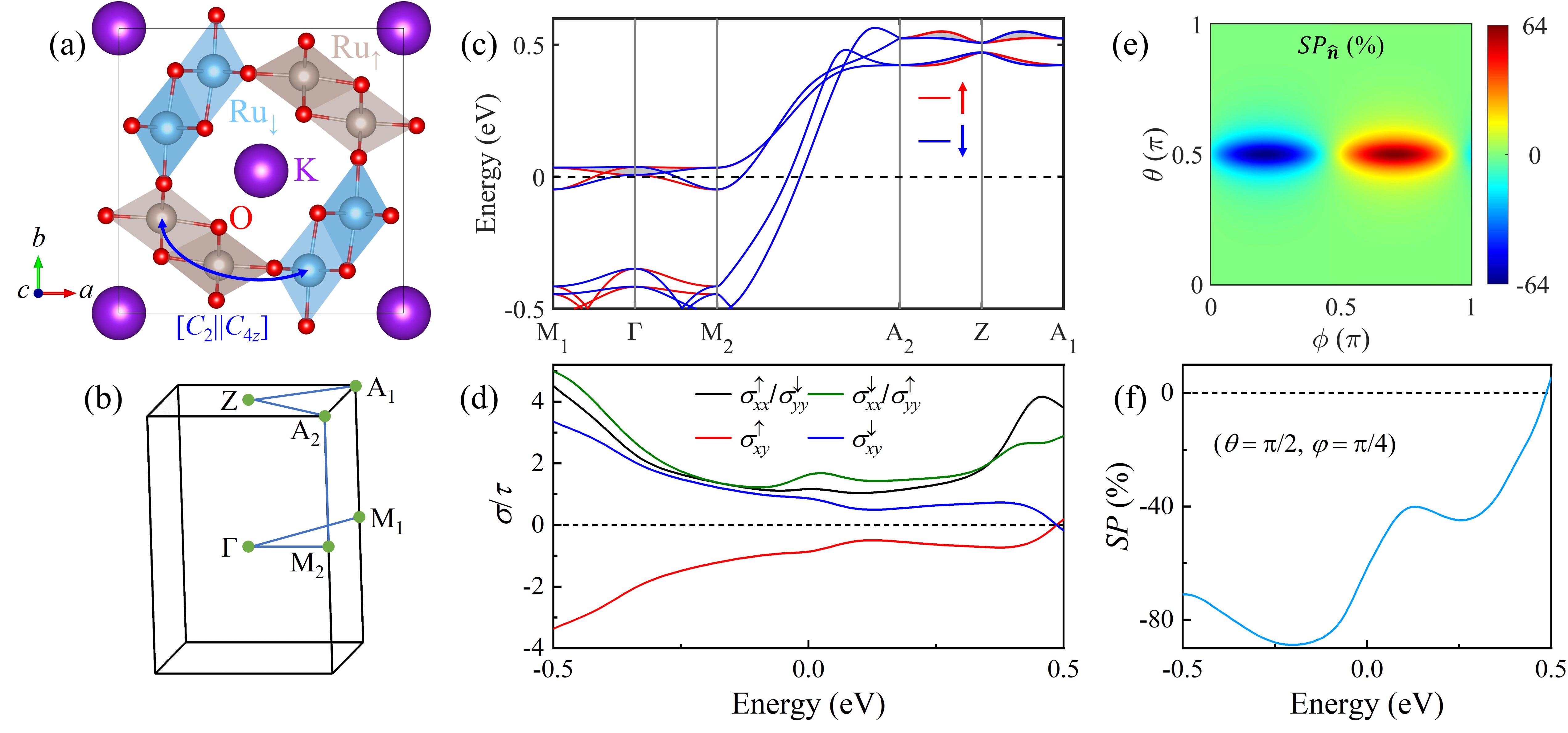}%
\caption{\label{f-2} (a) Crystal structure and (b) Brillouin zone of the tetragonal K$_2$Ru$_8$O$_{16}$ (space group $I4/m$). In (a), the opposite-spin sublattices are connected by the symmetry $[C_2||C_{4z}]$. (c) Spin-resolved band structure (Red for spin up and blue for spin down) without SOC along the $k$-space path: $M_1(-0.5, 0.5, 0)-\Gamma(0, 0, 0)-M_2(0.5, 0.5, 0)-A_2(0.5, 0.5, 0.5)-Z-A_1(-0.5, 0.5, 0.5)$ indicated in (b). The gray colored regions highlight the spin splitting. (d) Spin-resolved conductivity (unit: $10^{18}$ $\Omega^{-1}$m$^{-1}$s$^{-1}$) as a function of energy. The Fermi energy has been aligned to zero. (e) Spin polarization $SP_\mathbf{\hat{n}}$ at the Fermi energy as a function of ($\theta$, $\varphi$). (f) Spin polarization $SP$ at ($\theta=\pi/2$, $\varphi=\pi/4$) as a function of energy.}
\end{figure*}

K$_2$Ru$_8$O$_{16}$ crystallizes in a body centered tetragonal structure with the space group $I4/m$ shown in Fig. \ref{f-2}(a) and Fig. \ref{f-2}(b) shows its Brillouin zone with high-symmetry $k$ points indicated. From Fig. \ref{f-2}(a), the opposite-spin sublattices are connected by the symmetry $[C_2||C_{4z}]$ and the corresponding SPG is $^24/^{1}m$. We used the experimental crystal structure\cite{jlcm323,prb195101} of K$_2$Ru$_8$O$_{16}$ with lattice constants $a=9.866$, $c=3.131$ {\AA} and Monkhorst-Pack mesh of $4\times4\times12$ for DFT calculations. A Hubbard-$U$ correction of $U_{eff}=3.0$ eV on the Ru-$4d$ orbital was adopted to capture the electron correlations\cite{prb195101}. Figure \ref{f-2}(c) shows the spin-resolved band structure without SOC. We find that the nonrelativistic spin splitting appears along the $\Gamma-M_{1,2}$ and $Z-A_{1,2}$ paths. This is expected since the $k$ points along $\Gamma-M_{1,2}$ and $Z-A_{1,2}$ are not invariant under the symmetry operations in the subgroup $\mathbf{G}-\mathbf{H}=\{C_{4z}, C^3_{4z}, S_{4z}, S^3_{4z}\}$ while the $k$ points along $M_{2}-A_2$ remain invariant.

Figure \ref{f-2}(d) shows the spin-resolved conductivity as a function of energy calculated by using the $k$-point mesh of $60\times60\times200$. We see that $\sigma_{xx}$, $\sigma_{yy}$ and $\sigma_{xy}$ are spin polarized characterized by $\sigma^\uparrow_{xx}=\sigma^\downarrow_{yy}$, $\sigma^\uparrow_{yy}=\sigma^\downarrow_{xx}$ and $\sigma^\uparrow_{xy}=-\sigma^\downarrow_{xy}$ as consistent with the above SPG symmetry analysis (see Table \ref{table1}). From Eqs. \ref{eq-6} and \ref{eq-7} and Table \ref{table1}, the anisotropic spin polarization $SP_\mathbf{\hat{n}}$ can be calculated as
\begin{equation}\label{eq-9}
  SP_\mathbf{\hat{n}}=\frac{(\sigma_{xx}^\uparrow-\sigma_{xx}^\downarrow)\text{sin}^2\theta\text{cos}(2\varphi)+2\sigma_{xy}^\uparrow\text{sin}^2\theta\text{sin}(2\varphi)}{\sigma_{xx}\text{sin}^2\theta+\sigma_{zz}\text{cos}^2\theta}.
\end{equation}
Equation \ref{eq-9} suggests that $SP_\mathbf{\hat{n}}$ reveals a period $\pi$ for $\varphi$ and reaches extrema at $\theta=\pi/2$. In Fig. \ref{f-2}(e), we plot the $SP_\mathbf{\hat{n}}$ at the Fermi energy as a function of ($\theta$, $\varphi$). Again, $SP_\mathbf{\hat{n}}$ reveals significant spatial anisotropy and the maximum magnitude is more than $60\%$. In addition, $SP_\mathbf{\hat{n}}$ decays faster when $|\theta-\pi/2|$ increases for a fixed $\phi$ as compared to Fig. \ref{f-1}(e). Figure \ref{f-2}(f) shows $SP_\mathbf{\hat{n}}$ at ($\theta=\pi/2$, $\varphi=\pi/4$) as a function of energy. It is seen that the magnitude of $SP_\mathbf{\hat{n}}$ maintains $\sim60\%$ around the Fermi energy.

\subsection{RuO$_2$}

\begin{figure*}
\includegraphics[width=0.8\textwidth]{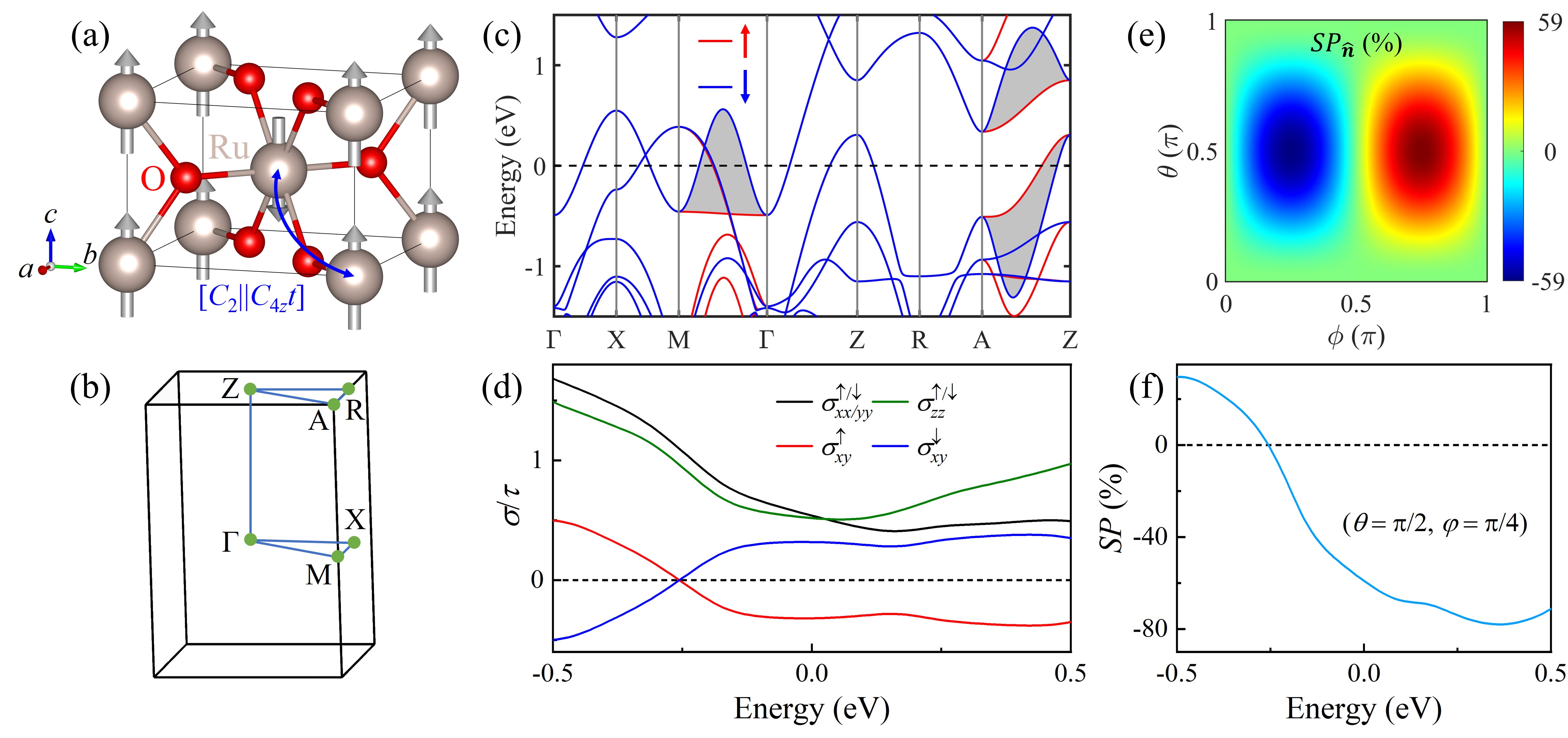}%
\caption{\label{f-3} (a) Crystal structure and (b) Brillouin zone of the tetragonal RuO$_2$ (space group $P4_2/mnm$). In (a), the gray arrows denote the magnetic moments of Ru atoms and the opposite-spin sublattices are connected by the symmetry $[C_2||C_{4z}t]$ with the translation $t=(1/2, 1/2, 1/2)$. (c) Spin-resolved band structure (Red for spin up and blue for spin down) without SOC along the $k$-space path: $\Gamma(0, 0, 0)-X(0, 0.5, 0)-M(0.5, 0.5, 0)-\Gamma-Z(0, 0, 0.5)-R(0, 0.5, 0.5)-A(0.5, 0.5, 0.5)-Z$ indicated in (b). The gray colored regions highlight the spin splitting. (d) Spin-resolved conductivity (unit: $10^{20}$ $\Omega^{-1}$m$^{-1}$s$^{-1}$) as a function of energy. The Fermi energy has been aligned to zero. (e) Spin polarization $SP_\mathbf{\hat{n}}$ at the Fermi energy as a function of ($\theta$, $\varphi$). (f) Spin polarization $SP$ at ($\theta=\pi/2$, $\varphi=\pi/4$) as a function of energy.}
\end{figure*}

RuO$_2$ crystallizes in a rutile structure with the space group $P4_2/mnm$ shown in Fig. \ref{f-3}(a) and Fig. \ref{f-3}(b) shows its Brillouin zone with high-symmetry $k$ points indicated. From Fig. \ref{f-3}(a), the opposite-spin sublattices are connected by the symmetry $[C_2||C_{4z}t]$ with the translation $t=(1/2, 1/2, 1/2)$ and the corresponding SPG is $^24/^1m^2m^1m$. We used the experimental crystal structure\cite{prl077201} of RuO$_2$ with lattice constants $a=4.492$ and $c=3.106$ {\AA} and Monkhorst-Pack mesh of $8\times8\times12$ for DFT calculations. A Hubbard-$U$ correction of $U_{eff}=3.0$ eV on the Ru-$4d$ orbital was adopted to capture the electron correlations. The symmetry elements $[C_2||M_{x}]$ and $[C_2||M_{y}]$ enforce the spin-degenerate nodal planes at $k_x=0, \pm\pi/a$ and $k_y=0, \pm\pi/b$, respectively. This is confirmed by the spin-resolved band structure shown in Fig. \ref{f-3}(c) and the sizable band splitting can be seen along $\Gamma-M$ and $A-Z$ paths consistent with previous work\cite{prx031042,prx040501}.

Figure \ref{f-3}(d) shows the spin-resolved conductivity as a function of energy calculated by using the $k$-point mesh of $140\times140\times200$. It is observed that only $\sigma_{xy}$ is spin polarized as characterized by $\sigma^\uparrow_{xy}=-\sigma^\downarrow_{xy}$ in accordance with the above SPG symmetry analysis (see Table \ref{table1}). From Eqs. \ref{eq-6} and \ref{eq-7} and Table \ref{table1}, the anisotropic spin polarization $SP_\mathbf{\hat{n}}$ can be calculated as
\begin{equation}\label{eq-10}
  SP_\mathbf{\hat{n}}=\frac{2\sigma_{xy}^\uparrow\text{sin}^2\theta\text{sin}(2\varphi)}{\sigma_{xx}\text{sin}^2\theta+\sigma_{zz}\text{cos}^2\theta}.
\end{equation}
Similarly, $SP_\mathbf{\hat{n}}$ reveals a period $\pi$ for $\varphi$ and reaches extrema at $\theta=\pi/2$. Figure \ref{f-3}(e) shows $SP_\mathbf{\hat{n}}$ at the Fermi energy as a function of ($\theta$, $\varphi$), which is quite similar to that observed for FeSb$_2$ [see Fig. \ref{f-1}(e)]. As can be seen, the large magnitude of $SP_\mathbf{\hat{n}}$ centers around ($\theta=\pi/2$, $\varphi=\pi/4$) and ($\theta=\pi/2$, $\varphi=3\pi/4$), where a large spin polarization of $\sim60\%$ can be obtained. Figure \ref{f-3}(f) shows $SP_\mathbf{\hat{n}}$ at ($\theta=\pi/2$, $\varphi=\pi/4$) as a function of energy. It reveals that the magnitude of spin polarization increases monotonically with increasing energy from the crossing point. A large spin polarization of $\sim60\%$ maintains around the Fermi energy.

\subsection{CuF$_2$}

\begin{figure*}
\includegraphics[width=0.8\textwidth]{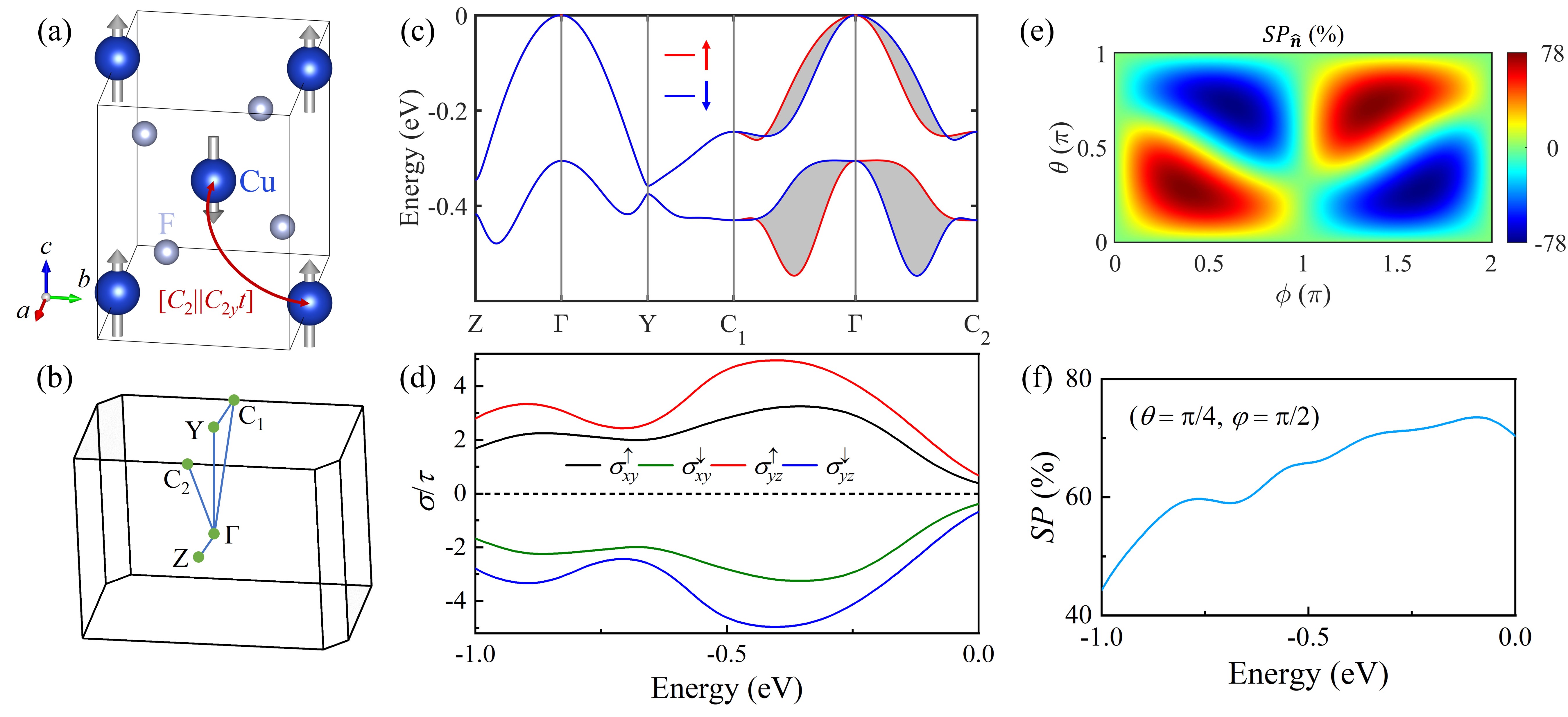}%
\caption{\label{f-4} (a) Crystal structure and (b) Brillouin zone of the monoclinic CuF$_2$ (space group $P2_1/c$). In (a), the gray arrows denote the magnetic moments of Cu atoms and the opposite-spin sublattices are connected by the symmetry $[C_2||C_{2y}t]$ with the translation $t=(0, 1/2, 1/2)$. (c) Spin-resolved band structure (Red for spin up and blue for spin down) without SOC along the $k$-space path: $Z(0, 0, 0.5)-\Gamma(0, 0, 0)-Y(0, 0.5, 0)-C_1(0, 0.5, -0.5)-\Gamma-C_2(0, 0.5, 0.5)$ indicated in (b). The gray colored regions highlight the spin splitting and the valence band maximum has been aligned to zero. (d) Spin-resolved conductivity (unit: $10^{19}$ $\Omega^{-1}$m$^{-1}$s$^{-1}$) as a function of energy. (e) Spin polarization $SP_\mathbf{\hat{n}}$ at $-0.1$ eV as a function of ($\theta$, $\varphi$). (f) Spin polarization $SP$ at ($\theta=\pi/4$, $\varphi=\pi/2$) as a function of energy.}
\end{figure*}

CuF$_2$ crystallizes in a monoclinic structure with the space group $P2_1/c$ shown in Fig. \ref{f-4}(a) and Fig. \ref{f-4}(b) shows its Brillouin zone with high-symmetry $k$ points indicated. We used the experimental lattice parameters\cite{jap1167} $a=3.30$, $b=4.57$ and $c=5.366$ {\AA} ($\beta=121.15^\text{o}$) and Monkhorst-Pack mesh of $12\times10\times8$ for DFT calculations. A Hubbard-$U$ correction of $U_{eff}=4.0$ eV on the Cu-$d$ orbital was adopted to capture the Mott insulator\cite{ssc1703}. As shown in Fig. \ref{f-4}(a), the opposite-spin sublattices are connected by the symmetry $[C_2||C_{2y}t]$ with the translation $t=(0, 1/2, 1/2)$ and the corresponding SPG is $^22/^2m$. Figure \ref{f-4}(c) shows the spin-resolved band structure without SOC for the valence bands. It is seen that the bands are spin degenerate along the $\Gamma-Z$ and $Y-C_1$ paths as protected by the symmetry operation $[C_2||M_y]$ and along the $\Gamma-Y$ path as protected by the symmetry operation $[C_2||C_{2y}]$. A sizable band splitting occurs along the $C_1-\Gamma-C_2$ path.

Figure \ref{f-4}(d) shows the spin-resolved conductivity as a function of energy around the top of valence bands as calculated by using the $k$-point mesh of $200\times140\times120$. We see that $\sigma_{xy}$ and $\sigma_{yz}$ are spin polarized as characterized by $\sigma^\uparrow_{xy}=-\sigma^\downarrow_{xy}$ and $\sigma^\uparrow_{yz}=-\sigma^\downarrow_{yz}$ in accordance with the above SPG symmetry analysis (see Table \ref{table1}). From Eqs. \ref{eq-6} and \ref{eq-7} and Table \ref{table1}, the anisotropic spin polarization $SP_\mathbf{\hat{n}}$ can be calculated as
\begin{widetext}
\begin{equation}\label{eq-11}
  SP_\mathbf{\hat{n}}=\frac{2\sigma_{xy}^\uparrow\text{sin}^2\theta\text{sin}(2\varphi)+2\sigma^\uparrow_{yz}\text{sin}(2\theta)\text{sin}\varphi}{\sigma_{xx}\text{sin}^2\theta\text{cos}^2\varphi+\sigma_{yy}\text{sin}^2\theta\text{sin}^2\varphi+\sigma_{zz}\text{cos}^2\theta+\sigma_{xz}\text{sin}(2\theta)\text{cos}\varphi}.
\end{equation}
\end{widetext}
$SP_\mathbf{\hat{n}}$ reveals a period $2\pi$ for $\varphi$, which is distinct to the $SP_\mathbf{\hat{n}}$ demonstrated for the above SPGs $^2m^2m^1m$, $^24/^{1}m$ and $^24/^1m^2m^1m$. In addition, as show in Fig. \ref{f-4}(e), $SP_\mathbf{\hat{n}}$ reveals a more intricate ($\theta$, $\varphi$) phase diagram due to the distinct ($\theta$, $\varphi$) dependencies of $\sigma_{xy}^\uparrow$ and $\sigma^\uparrow_{yz}$ in the numerator of Eq. \ref{eq-11}. Moreover, the maximum magnitude of $SP_\mathbf{\hat{n}}$ at the Fermi energy reaches nearly $80\%$. Figure \ref{f-4}(f) shows $SP_\mathbf{\hat{n}}$ at ($\theta=\pi/4$, $\varphi=\pi/2$) as a function of energy. As can be seen, a large spin polarization of $\sim70\%$ maintains around the top of valence bands.

\subsection{CrSb}
\begin{figure}
\includegraphics[width=0.45\textwidth]{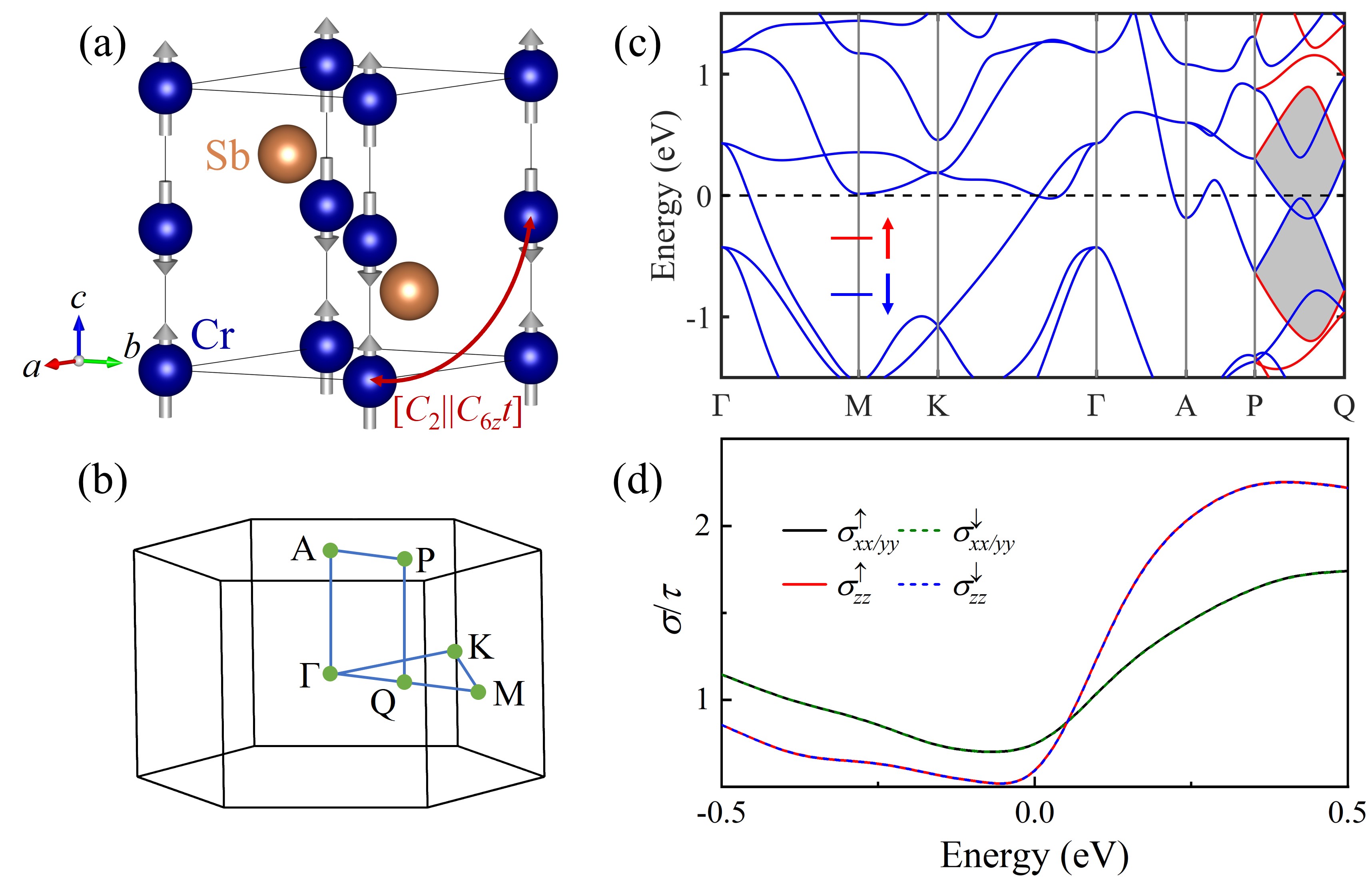}%
\caption{\label{f-5} (a) Crystal structure and (b) Brillouin zone of the hexagonal CrSb (space group $P6_3/mmc$). In (a), the gray arrows denote the magnetic moments of Cr atoms and the opposite-spin sublattices are connected by the symmetry $[C_2||C_{6z}t]$ with the translation $t=(0, 0, 1/2)$. (c) Spin-resolved band structure (Red for spin up and blue for spin down) without SOC along the $k$-space path: $\Gamma(0, 0, 0)-M(0.5, 0, 0)-K(1/3, 1/3, 0)-\Gamma-A(0, 0, 0.5)-P(0.25, 0, 0.5)-Q(0.25, 0, 0)$ indicated in (b). The gray colored regions highlight the spin splitting. The Fermi energy has been aligned to zero. (d) Spin-resolved conductivity (unit: $10^{20}$ $\Omega^{-1}$m$^{-1}$s$^{-1}$) as a function of energy.}
\end{figure}

CrSb crystallizes in a hexagonal NiAs-type structure with the space group $P6_3/mmc$ shown in Fig. \ref{f-5}(a) and Fig. \ref{f-5}(b) shows its Brillouin zone with high-symmetry $k$ points indicated. We used the experimental crystal structure\cite{prl206401} of CrSb with lattice constants $a=4.124$ and $c=5.473$ {\AA} and Monkhorst-Pack mesh of $12\times12\times10$ for DFT calculations. From Fig. \ref{f-5}(a), the opposite-spin sublattices are connected by the symmetry $[C_2||C_{6z}t]$ with the translation $t=(0, 0, 1/2)$ or the symmetry $[C_2||M_{z}]$ and the corresponding SPG is $^26/^2m^2m^1m$. In particular, the symmetry elements $[C_2||C_{6z}t]$ and $[C_2||M_{z}]$ enforce four spin-degenerate nodal planes in the Brillouin zone\cite{prx031042,prl206401}. Figure \ref{f-5}(c) shows the spin-resolved band structure without SOC. It is seen that the bands are spin degenerate along high-symmetry $k$-space paths on those nodal planes. In contrast, significant band splitting occurs along the low-symmetry path $P-Q$ consistent with previous work\cite{nc2116}.

Figure \ref{f-5}(d) shows the spin-resolved conductivity as a function of the energy calculated by using the $k$-point mesh of $200\times200\times150$. It is seen clearly that all the conductivity components are spin degenerate as consistent with SPG symmetry analysis [see Table \ref{table1}]. Thus, the anisotropic spin polarization is zero in CrSb.

\section{Discussion and summary}

In this work, we calculate the spin-polarized conductivity in altermagnets based on the relaxation time $\tau$ approximation. In general, the scattering probability is spin and wave vector dependent, which calls for the advanced Boltzmann transport theory beyond $\tau$ approximation\cite{pbr035436}. Since the spin polarization in altermagnets is dominated by the intrinsic band splitting, we expect that the Boltzmann transport theory beyond $\tau$ approximation does not affect the qualitative results, but influences the quantitative outcomes. Second, the collinear altermagnets are also demonstrated in two-dimensional systems\cite{apl182409,cs13853,jcas381}. It is thus interesting to explore the effect of dimensionality on the anisotropic spin polarization in the future study. Third, it is worth mentioning that the polarity of spin splitting is locked to the direction of the N\'eel vector in altermagnets. Thus, it is expected that the spin polarization can be reversed by the the N\'eel vector reversal in altermagnets due to the swap between spin up and down channels. Finally, it is shown that the concept of altermagnet can be extended to the antiferromagnet with non-collinear spins\cite{npj13}, which calls for the further exploration on the spin polarization in the non-collinear altermagnet.

In summary, using the Boltzmann transport theory, we have investigated the anisotropic spin polarization in collinear altermagnets based on the group-theoretical analysis and DFT calculations. We show that only the $d$-wave altermagnet sustains the finite spin polarization, which reveals significant spatial anisotropy. In particular, we derive the analytical expressions of the anisotropic spin polarization for different spin point groups. Our results are expected to enrich the altermagnetic physics and provide a design strategy to produce large spin polarization in altermagnets.

\begin{center}
{\bf ACKNOWLEDGMENTS}
\end{center}

This research was supported by the National Natural Science Foundation of China (Grant No. 12274102) and the Fundamental Research Funds for the Central Universities (Grant No. FRFCU5710053421, No. HIT.OCEF.2023031). The atomic structures were produced using the VESTA software\cite{vesta}.

\begin{center}
{\bf DATA AVAILABILITY}
\end{center}

The data that support the findings of this article are not publicly available upon publication because it is not technically feasible and/or the cost of preparing, depositing, and hosting the data would be prohibitive within the terms of this research project. The data are available from the authors upon reasonable request.

\end{document}